\documentclass[prl,twocolumn,showpacs,floatfix,amsfonts]{revtex4}
\usepackage{graphicx,graphics,color,epsfig} 
\usepackage{bm}
\usepackage{amsmath}
\usepackage{amssymb}
\begin{document}

\title{Spin-Orbit Coupling and Tunneling Current in 
a Parabolic Quantum Dot}
\author{Hong-Yi Chen$^*$, Vadim Apalkov$^\dagger$, and Tapash 
Chakraborty$^*$}
\affiliation{$^*$Department of Physics and Astronomy, University of 
Manitoba, Winnipeg, MB R2T 2N2, Canada \\
$^\dagger$Department of Physics and Astronomy, Georgia State University, 
Atlanta, Georgia 30303, USA}

\begin{abstract}
We propose a novel approach to explore the properties of a quantum dot 
in the presence of the spin-orbit interaction and in a tilted magnetic 
field. The spin-orbit coupling within the quantum dot manifest itself 
as anti-crossing of the energy levels when the tilt angle is varied. 
The anti-crossing gap has a non-monotonic dependence on the magnitude 
of the magnetic field and exhibits a peak at some finite values of 
the magnetic field. From the dependence of the tunneling current through 
the quantum dot on the bias voltage and the tilt angle, the anti-crossing 
gap and most importantly the spin-orbit coupling strength can be uniquely 
determined. 
\end{abstract}

\maketitle

In recent years, there has been a well concerted effort to achieve a
coherent control on the electron spin transport in semiconductor 
nanostructures because of its attractive potential for future spin-based 
electronic devices \cite{spintro,ohno}. The spin-orbit (SO) interaction 
plays a crucial role in that pursuit as it provides a means for coupling 
of the electron spin to its orbital motion. The SO interaction may in 
turn be manipulated by applying a gate voltage. Studies of the SO 
coupling effects in parabolic quantum dots are equally intriguing 
\cite{tc,pekka} because it is expected that such a system will provide 
the important step toward the quantum information processing \cite{loss}. 
In narrow-gap semiconductors such as the InAs-based systems, the dominant 
source of the SO interaction is the structural inversion asymmetry 
\cite{sia}. The resultant Bychkov-Rashba type of SO interaction \cite{bychkov} 
is the interaction of our choice in the present investigation. 
The most common method of determining the strength of the SO coupling 
is to study the beating pattern in Shubnikov-de Haas (SdH) 
oscillations \cite{expt}. However, that process does not always provide an 
unambiguous determination of the SO coupling strength. 

In this letter, we propose a new theoretical approach for measurement 
of the strength of the SO interaction in quantum dots. This approach is 
based on an analysis of the behavior of the electronic quantum dot energy 
levels in a tilted magnetic field. The tilted magnetic field has the 
distinct advantage over parallel and perpendicular fields because 
it introduces the Zeeman splitting of the energy levels and modifies 
the orbital motion of the electron within the quantum dot as well. The 
relative strength of these two contributions in the electron dynamics 
can be varied by changing the tilt angle. Without 
the SO interaction the energy spectrum of the quantum dot has a strong 
dependence on the direction of the magnetic field exhibiting regions 
of level crossings at different tilt angles. The levels that cross 
have the opposite spin directions, and without the SO interaction 
there is no mixing between them. Introducing the SO interaction results 
in a coupling between the different spin states. In this case we 
should expect an anti-crossing of the energy levels as a function of 
the tilt angle. The strength of the anti-crossing characterizes the 
strength of the SO coupling. The most accurate way to study experimentally 
the structure of the energy spectra around the anti-crossing region is 
to measure the tunneling current through the quantum dot system. 
Transport spectroscopy is a powerfull tool to study a variety of phenomena 
related to the correlation and interaction effects in a quantum dot 
\cite{transportQD,anisotropy}. The main idea of the tunneling spectrosopy 
at a finite bias voltage is that the tunneling current depends on the 
number of available (for tunneling) channels in the quantum dot. In the 
following, we study the tunneling transport through a quantum dot in 
a tilted magnetic field and show that the tunneling current has a 
unique dependence on the tilt angle and the bias voltage within the 
anti-crossing region. 

The energy range of the anti-crossing region is usually smaller than the 
energy of the inter-electron interaction, which can be estimated to be about
7 meV \cite{pekka}. In that case we can describe the tunneling process by 
a single-electron picture. The Hamiltonian of an electron 
in a parabolic quantum dot in a tilted magnetic field has the form 
\begin{eqnarray}
\nonumber
H &=& \frac1{2m^*} \left( {\bf p} + \frac ec{\bf A} \right)^2
  + \frac12 m^* \omega_0^2 r^2 + \frac12 g \mu_B B_z \sigma_z \\
\nonumber
&&+ \frac{\alpha}{\hbar} \left[ {\bf \sigma} \times \left( {\bf p} +
  \frac{e}{c}{\bf A} \right) \right]_z
  + \frac12 g \mu_B B_x \sigma_x \;.
\end{eqnarray}
Here, ${\bf A}=\frac12B_z(-y,x,0)$ is the vector potential in the
symmetric gauge, $\alpha$ is the spin-orbital coupling strength, $g$ 
is the effective Land\'e $g$ factor, and $\bf p$ is the two-dimensional
vector in the $(x,y)$ plane. In the above equation we assumed that there is 
no dynamics in the $z$ direction due to the size quantization and the electron 
occupies the corresponding lowest subband. The value of $\alpha$ obtained from 
various experiments lie in the range of 5 -- 45 meV$\cdot$nm \cite{expt}.
In a tilted magnetic field, the perpendicular component is $B_z=B \cos \theta$ 
while the parallel component is $B_x=B \sin \theta$, where $B$ is the 
magnitude of magnetic field and $\theta$ is the angle between the magnetic 
field vector and the $z$-axis. In the above expression for the vector potential 
${\bf A}$, we have taken into account only the perpendicular component of 
the magnetic field $B_z$. Since the size of the dot in the $z$ direction is 
small, the only effect of the parallel magnetic field is through the Zeeman 
energy. The energy spectra and the wavefunctions corresponding to the 
above Hamiltonian (but for a zero tilt angle) have been obtained earlier 
numerically \cite{pekka}. All the calculations below have been performed 
for the case of InAs quantum dots. It should be pointed out that titled-field
experiments on the quantum dots have been reported earlier in the literature
\cite{tilted_expt}, but in the absence of the SO coupling. 

In our approach, a quantum dot is attached through the tunneling barriers 
to the right and left leads. We study the tunneling current through the dot 
at a finite bias voltage between the leads. We describe the process of 
tunneling through a parabolic quantum dot as a sequential single-electron 
tunneling \cite{seqTun}. The quantum dot system is charactrized by the 
probability $P_0$ that there are no electrons in the dot and probabilities 
$P_i, i=1,\cdots,N$ that the electron occupy an energy level $E_i$ in the 
dot. For the probabilitiy $P_i$ we can write the rate equations in the form 
\begin{eqnarray}
&& \frac{\partial P_0}{\partial t} = -P_0 \sum_{i=1}^N W_i + 
\sum_{i=1}^N P_i V_i \;, {}\\
&& \frac{\partial P_i}{\partial t} = -P_i V_i + W_i P_0 \;, {}\\
&& P_0 + P_1 + P_2 + \cdots + P_N = 1 \;,
\end{eqnarray}
where the last equation is the normalization condition. Here the 
transition rates $W_i$ and $V_i$ are the rates of tunneling in and 
out of the dot, respectively. These rates can be found from the 
Fermi golden rule 
\begin{eqnarray*}
&& W_i = \Gamma f_L(E_i) + \Gamma f_R(E_i) \;,\\
&& V_i = \Gamma (1-f_L(E_i)) + \Gamma (1-f_R(E_i)) \;,
\end{eqnarray*}
where $\Gamma$ is the tunneling rate, which we assume to be energy 
independent and is also the same for both left and right leads. Here $f_L(E)$ 
and $f_R(E)$ are the Fermi distribution functions of the left $(L)$ and rigth 
$(R)$ leads, respectively. The chemical potentials of the left and right leads are 
$\mu_L$ and $\mu_R$ respectively. In the calculations that follow, we have chosen the 
the ground state of a quantum dot with a single electron as the zero-energy 
state. The temperature in our calculation is $\sim 10$K.

For the stationary case, the time derivatives of $P_0$ and $P_i$ 
are zero. Then the linear system of equations Eqs.~(1)-(3) can be 
easily solved and the stationary tunneling current can be found 
from the equation 
\begin{eqnarray*}
I(V) = \sum_{i=1}^N \left( W_i^L P_0 - V_i^L P_i \right) \;,
\end{eqnarray*}
where $V$ is the bias voltage and the chemical potentials $\mu_L$ 
and $\mu_R$ are related to $V$ as $\mu_L=V/2$ and $\mu_R=-V/2$.
 
\begin{figure}[!ht]
\centerline{\epsfxsize=8.5cm\epsfbox{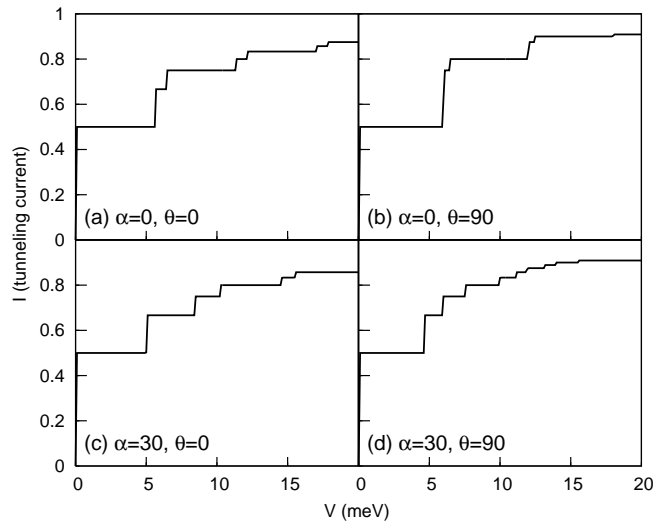}}
\caption[*] {
Tunneling current vs the bias voltage for four different cases 
at $B=4$ Tesla: (a) $\alpha=0, \theta=0$; (b) $\alpha=0$, 
$\theta=90^\circ$; (c) $\alpha=30$ meV$\cdot$nm, $\theta=0$; 
(d) $\alpha=30$ meV$\cdot$nm, $\theta=90^\circ$. The parameters for 
InAs quantum dots are $m^*/m_0=0.042$, $g=-14$, and the confinement 
potential strength is $\hbar\omega_0=3.0$ meV.
}
\end{figure}

In Fig.~1, we show the tunneling current as a function of the bias voltage 
for four different cases.  The four cases can be divided into two groups 
by the angle of the applied tilted magnetic field: (i) ($\theta=0^\circ$) 
and (ii) ($\theta=90^\circ$). In the first case, Fig.~1(a) and Fig.~1(c) 
do not show any significant difference when the SO interaction is included, 
while in the second case the presence of the SO interaction lifts the 
degenerate states which creates more steps in the I-V curve [as seen in 
Fig.~1(b) and Fig.~1(d)].

\begin{figure}[!ht]
\centerline{\epsfxsize=7.0cm\epsfbox{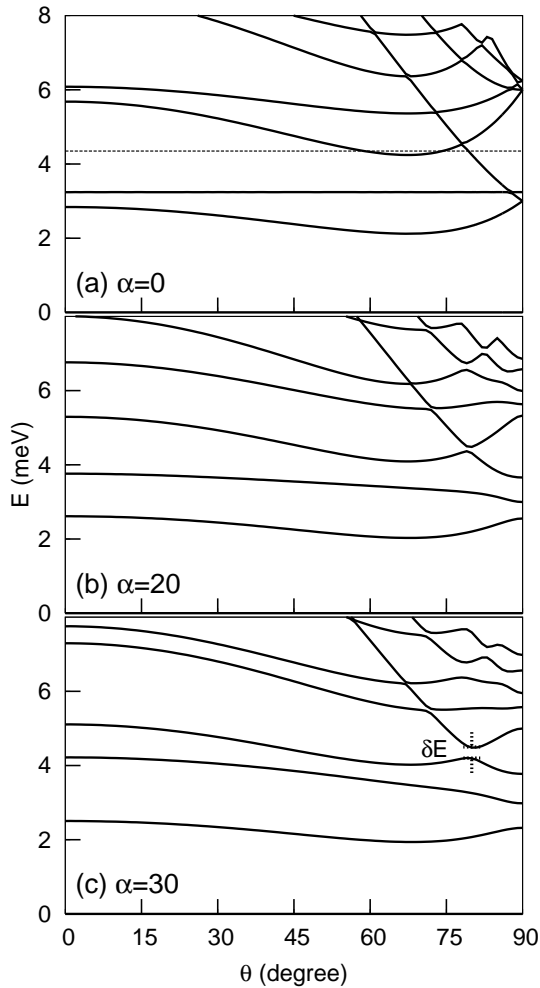}}
\caption[*] {The energy spectra as a function of the tilt angle 
($\theta$) for $B=4$ Tesla and for different values of the SO 
coupling strength: (a) $\alpha=0$, (b) $\alpha=20$, and (c) $\alpha=30$  
meV$\cdot$nm.  The dashed line in (a) corresponds to the energy $E=4.3$ 
meV. In (c), $\delta E$ is the energy gap.}
\end{figure}

From Fig.~1 it is clear that by varying the tilt angle $\theta$ one 
can make a significant change in the I-V curve. In order to study the 
effect of a tilted field, we have looked at the angle dependence of 
the energy levels. Figure~2(a) shows several level crossing in the absence 
of the SO coupling. The first crossing appears around $E=4.5$ meV and 
$\theta$ between 70$^\circ$ and 90$^\circ$. In the presence of the SO 
coupling [Fig.~2(b)], that level crossing becomes an anti-crossing with an 
energy gap of $\delta E$. Figure~2(c) shows that the energy gap increases 
with an increase of the strength of the SO coupling. The anti-crossing in 
Fig.~2 is a direct manifestation of the SO interaction. In what follows, 
we demonstrate that the anti-crossing of the energy levels results in a 
specific dependence of the tunneling current on the bias voltage and the 
tilt angle. 

\begin{figure}[!ht]
\centerline{\epsfxsize=7.0cm\epsfbox{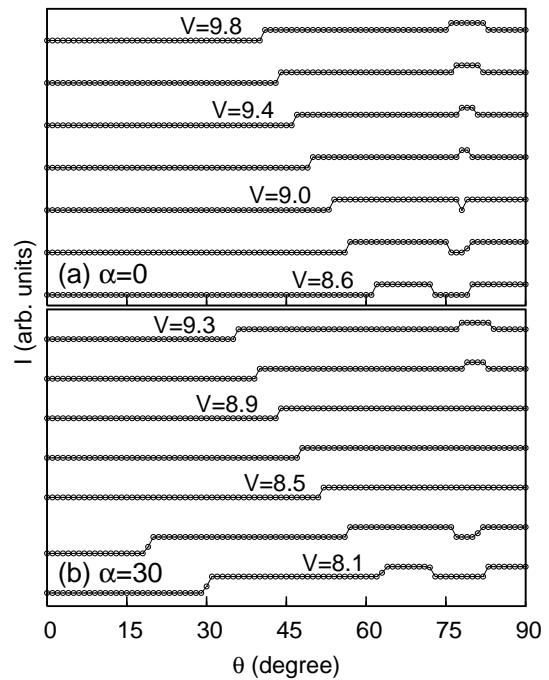}}
\caption[*] {Tunneling current as a function of the tilt angle $\theta$ 
at $B=4$ Tesla and for (a) $\alpha=0$, and (b) $\alpha=30$ meV$\cdot$nm.  
Each line corresponds to a constant bias voltage $V$. The bias voltage is 
expressed in meV. The increment of the voltage is 0.2 meV. The curves have 
been shifted vertically for clarity.
}
\end{figure}

The tunneling current as a function of $\theta$ is shown in Fig.~3.  
In Fig.~3(a) we present the data for the tunneling current at different 
bias voltages with an increment of 0.2 meV for the quantum dot without a
SO coupling.  At $V=8.6$ meV ($\mu_L=4.3$ meV), the Fermi energy of the left 
lead $\mu_L$ is below the first level crossing, which is illustrated by 
the dashed line in Fig.~2(a). Around $\theta=70^\circ$, there are three 
levels of the quantum dot below $\mu_L$.  As we increase $\theta$ the 
Fermi energy of the left lead goes below the third energy level.  At 
this point, the tunneling current which depends on the number of levels 
between the Fermi energies of the left and right leads, drops. However, when 
$\theta \ge 80^\circ$, the Fermi energy $\mu_L$ is again above the third 
energy level. The tunneling current then goes up. As a result, the tunneling 
current as a function of the tilt angle shows a dip at the voltage below 
the crossing point. When we increase the voltage and approach the crossing 
point, the dip becomes narrower. Just above the crossing point, the tunneling 
current shows a narrow bump similar to that at $V=9.2$ meV in Fig~3(a).  
With a further increase of the bias voltage the bump in the tunneling current 
becomes wider.

Figure~3(b) shows the tunneling current for a finite value of the SO 
coupling strength $\alpha=30$  meV$\cdot$nm. Just as for the system 
without the SO interaction, the tunneling current reveals a dip when the 
bias voltage is less than $8.3$ meV.  When the voltage is increased from
that value, the system shows a behavior characteristic of that of the 
level anti-crossing. Namely, within a finite interval of the bias voltages 
$\delta V = 2 \delta E$, the tunneling current becomes independent of the 
tilt angle. This corresponds to the case where the Fermi energy of the 
left lead is in the anti-crossing gap. If the voltage is continuously 
increased, the flat pattern disappears and in its place a bump pattern 
emerges. Changing of the pattern reveals the evidence for the existence 
of the SO coupling which opens a gap at the crossing point [see Fig.~2(c)]. 
The dip occurs when the voltage is below the bottom edge of the energy gap, 
while the flat curve appears when the voltage is inside the gap. The 
bump in the curve means that the voltage is above the top edge of the 
energy gap. The change of pattern from a dip to being flat and then 
to a bump can be quantified by the voltage difference $\delta V$. 
Since $\delta V= 2\delta E$, this voltage difference will determine 
the strength of the SO coupling.

\begin{figure}[!ht]
\centerline{\epsfxsize=8.5cm\epsfbox{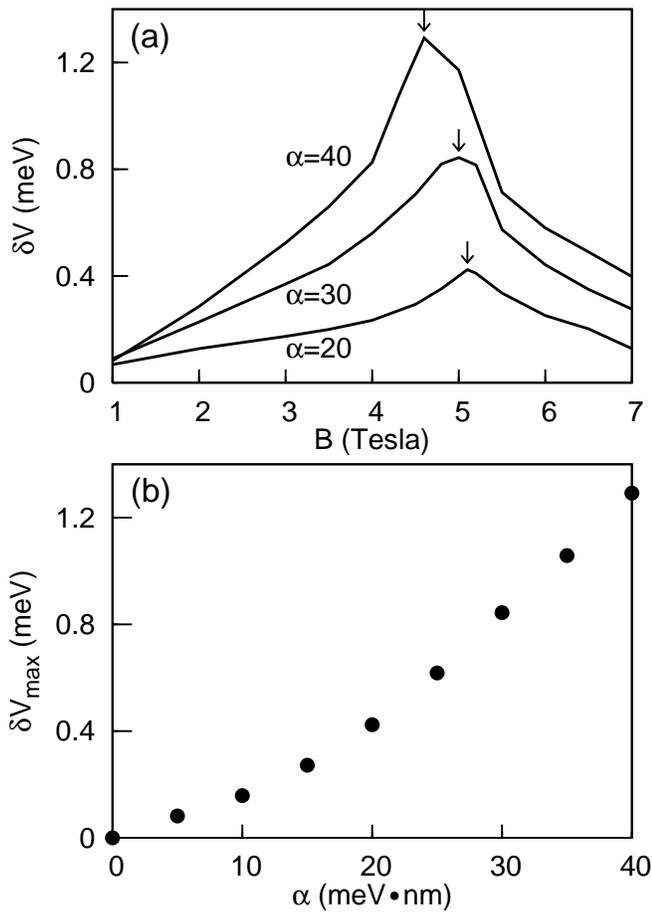}}
\caption[*] {(a) The magnetic field dependence of the voltage difference, 
$\delta V$ for three different values of the SO coupling strength: 
$\alpha=20$ 30, and $40$ meV$\cdot$nm. The corresponding peak positions 
are at $B=5.1$, $5.0$, and $4.6$ Tesla for $\alpha=20$, $30$, $40$, 
respectively. (b) The SO coupling strength dependence on the maximum 
voltage difference, $\delta V_{\rm max}$. Each point correspods to 
a different value of the magnetic field.}
\end{figure}

Analyzing the tunneling current as a function of the angles we are 
able to directly evaluate the strength of the SO coupling.  However, the 
anti-crossing energy gap depends not only on the SO coupling strength but 
also on the magnitude of the applied magnetic field. With an increasing 
magnetic field the size of the energy gap increases and reaches a maximum 
value $\delta E_{\rm max}=\delta V_{\rm max}/2$. Figure~4(a) illustrates 
the above trend for three different values of SO coupling strength. For 
larger values of the SO coupling strength, the peak is located at a lower 
magnetic field.  The peak shifts toward a higher field as the SO coupling 
strength decreases.  All the peak values are located between $B=4.5$ 
Tesla and $B=5.5$ Tesla. The optimal value of the magnetic field illustrates 
the interplay between the orbital and spin effects of the magnetic field. 
In Fig.~4(b) the value of $\delta V_{\rm max}$ at the optimal magnetic 
field is shown as a function of the SO coupling. Note that at different 
values of the SO coupling the optimal magnetic field is different in 
Fig.~4(b). With the known maximum value of the voltage differences, 
$\delta V_{\rm max}$, the corresponding SO coupling 
strength can be directly determined.


In conclusion, the energy spectra of a quantum dot system in a tilted 
magnetic field exhibits the anti-crossing behavior of the energy levels 
as a function of the tilt angle. The nature of anti-crossing of the 
energy levels is entirely due to the SO interaction. In the I-V 
characteristics of the tunneling current through the quantum dot the 
anti-crossing regions can be identified and the corresponding gap can 
be directly determined. The value of the gap has a strong dependence 
on the magnitude of the magnetic field and has a maximum at a finite 
value of the magnetic field. The anti-crossing gap exhibits a monotonic 
increase with an increase of the spin-orbit coupling strength.

HYC would like to thank P. Pietil\"ainen for helpful discussions. The 
work has been supported by the Canada Research 
Chair Program and a Canadian Foundation for Innovation Grant.

\end{document}